\documentclass[iop]{emulateapj}
\usepackage{graphicx}
\usepackage{amssymb}
\usepackage{epstopdf}
\usepackage{float}
\usepackage{verbatim}
\usepackage{natbib}
\usepackage{amsmath}
\usepackage{mathptmx}
\usepackage{float}
\usepackage[colorlinks=true,urlcolor=blue,citecolor=blue,linkcolor=blue]{hyperref}
\newcommand{\oi}{\ion{O}{1}}

\newcommand{\ci}{\ion{C}{1}}
\newcommand{\cii}{\ion{C}{2}}
\newcommand{\mgii}{\ion{Mg}{2}}
\newcommand{\oil}{\oi~135.56~nm line}
\newcommand{\cil}{\ci~135.58~nm line}
\newcommand{\iris}{\textit{IRIS}}
\newcommand{\eg}{\textit{e}.\textit{g}.}
\newcommand{\kms}{\text{km\,s}$^{-1}$}
\newcommand{\dd}{\ensuremath{\mathrm{d}}}
\def\edt#1{{#1}}

\shorttitle{THE FORMATION OF \iris\ DIAGNOSTICS}
\shortauthors{Hsiao-Hsuan Lin et al.}

\begin{document}
\title{THE FORMATION OF \iris\ DIAGNOSTICS \\
    IX. THE FORMATION OF THE C I 135.58 nm LINE IN THE SOLAR ATMOSPHERE}


\author{Hsiao-Hsuan Lin$^1$}
\author{Mats Carlsson\altaffilmark{1}}\email{mats.carlsson@astro.uio.no}
\author{Jorrit Leenaarts\altaffilmark{2,1}}\email{jorrit.leenaarts@astro.su.se}

\affil{$^1$ Institute of
 Theoretical Astrophysics, University of Oslo, P.O. Box 1029
 Blindern, NO--0315 Oslo, Norway}
\affil{$^2$ Institute for Solar Physics, Department of Astronomy, Stockholm University, AlbaNova University Centre, SE-106 91 Stockholm, Sweden}

\begin{abstract}
The \cil\ is located in the wavelength range of NASA's {\it Interface Region Imaging Spectrograph} (\iris) small explorer mission.  We here study the formation and diagnostic potential of this line by means of non local-thermodynamic-equilibrium
modeling, employing both 1D and 3D radiation-magnetohydrodynamic models. The 
 \ci/\cii\ ionization balance is strongly influenced by  photoionization by Ly$\alpha$ emission. The emission
in the \cil\ is dominated by a recombination cascade and the line forming region is optically thick.
The Doppler shift of the line correlates strongly with the vertical velocity in its line forming region, 
which is typically located at 1.5 Mm height. 
With \iris\ the \cil\ is usually observed together with the \oil, and from the Doppler shift of both lines, we obtain the velocity difference between the line forming regions of the two lines. 
From the ratio of the \ci/\oi\ line core intensity, we can determine the distance between the \ci\ and the \oi\ forming layers. Combined with the velocity difference, 
the velocity gradient at mid-chromospheric heights can be derived. 
The \ci/\oi\ total intensity line ratio is correlated with the inverse of the electron density in the mid-chromosphere. 
We conclude that the \cil\ is an excellent probe of the middle chromosphere by itself, and together with the \oil\ the two lines provide even more information, which complements other powerful 
chromospheric diagnostics of \iris\ such as the \ion{Mg}{2} h and k lines and the \cii\ lines around 133.5~nm.

\end{abstract}


\section{Introduction}

The \cil\ ($2s^2\,2p^2\,^1\!D_2$ - $ 2s^2\,2p\,4d\,^1\!F_3^o$) is in the wavelength range of the NASA small explorer
mission {\it Interface Region Imaging Spectrograph} ({\it IRIS}) and is normally observed together with the nearby
\oil\ \citep{hhlinoi}. Their line cores are formed in the mid-chromosphere, 
unlike other strong lines that \iris\ covers, such as \ion{Mg}{2} h and k \citep{mg1,mg2,mg3} and the \cii\ 133.5 nm multiplet 
\citep{bhavna1, bhavna2,bhavna3}, whose cores usually form in the upper chromosphere or transition region. 

Already,  the SKYLAB mission revealed interesting behavior of the \ci/\oi\ line ratio: during a solar flare, the
\cil\ gets stronger than the \oil, the opposite of the behavior in the quiet Sun \citep{OICI-cheng80}. 
It is thus clear that the \cil\ may provide diagnostics of the solar chromosphere, especially when combined with
observations in the \oil.

\cite{semiflareupper} modeled \ion{C}{1}-\ion{C}{4} with semi-empirical models of flares. 
The \ion{C}{1} 123.9 nm continuum is blended with the Ly$\alpha$ wing and the hydrogen radiative transfer 
was therefore solved together with  carbon to take into account the effect of
Ly$\alpha$ on the \ion{C}{1}/\ion{C}{2} ionization balance. It is  important to include 
effects of partial redistribution (PRD) in the Ly$\alpha$ line profile.
According to their calculations, the \ion{C}{1}/\ion{C}{2} ionization balance depends on the Ly$\alpha$ 
wing intensity and the Lyman continuum, but is insensitive to \ion{C}{1} line transfer.  
 
\citet{ci1561} presented a more complete study of the carbon line-formation problem, focussing on the \ci~156.1\,nm and 165.7\,nm lines, 
based on the VAL3C semi-empirical model \citep{VAL3}. The authors found that the upper levels of these transitions are
mostly populated by radiative recombination to highly excited levels, followed by radiative cascading. 
The central reversals in the calculated lines were deeper than in the observations. This could be due
to only including four levels with higher excitation potential than the upper levels of the studied transitions, thus
underestimating the radiative recombination, or it could be due to an underestimate of the radiative ionization
of \ion{C}{1} by hydrogen transitions.

\edt{Another} theoretical modeling of the non local-thermodynamic-equilibrium (non-LTE) neutral carbon 
spectral-line formation was made by \cite{fabbianci} for stellar abundance analysis. 
In their study, they focussed on \ion{C}{1} lines in the wavelength range of 830--960\,nm. These lines are located
both in the singlet and the triplet system of \ion{C}{1}. The authors found that radiative processes dominate over
collisional processes and the non-LTE effects are rather large. In metal-poor stars, also intersystem collisions between
the singlet and the triplet systems may also be important. For this reason, \citet{fabbianci} also included intersystem collisions
with hydrogen, following the Drawin recipe \citep{1968ZPhy..211..404D,1969ZPhy..225..483D} with a scaling factor.
They found that the resulting non-LTE corrections to abundance determinations using these \ion{C}{1} lines were
not very dependent on the neutral hydrogen collisions.

\edt{\citet{2008ApJS..175..229A} solve the non-LTE problem employing an extended \ci-\ion{C}{4} model atom with 128 individual energy levels
of \ci\ grouped into 19 reference levels. This is done together with non-LTE computations of hydrogen (15 bound levels), \oi-\ion{O}{6} 
(40 levels of \oi\ grouped into 20 reference levels), and 10 other atoms and ions. They produce a semi-empirical model of 
the quiet solar chromosphere and transition region that reproduces as closely as possible the detailed spectrum observed by SUMER 
\citep{2001A&A...375..591C}
with an additional constraint that the transition region is given by an energy equation containing radiative processes, conduction, and particle diffusion (in contrast with the VAL3C model 
\citep{VAL3}
where the transition region temperature structure was a fitting function as well).}

In this work, we will study the formation of the \cil\ in the solar chromosphere employing both a 1D semi-empirical model  
and a 3D radiation-magnetohydrodynamic (RMHD) model. We will also explore the diagnostic potential of combining the
observations in the \ci\ and \oi\ lines.   
In Section \ref{sec:method} we describe our model atmospheres, radiative transfer computations, and atomic models. 
In Section \ref{sec:steady} we discuss the basic formation mechanism and the effects from the hydrogen solution. 
Following that, we discuss how the dynamics in the atmosphere influences the formation mechanism in Section \ref{sec:move} 
and we present synthetic spectra calculated based on 3D realistic atmospheres. In Section \ref{sec:diag} we present their potential diagnostic value, and we conclude in Section \ref{sec:concl}.

 \section{Methods}\label{sec:method}
 
 \subsection{Radiative transfer computations}
In this study, we use the RH code \citep{RH-uitenbroek01} to solve the non-LTE radiative transfer problem. 
RH is a multilevel accelerated lambda iteration code for radiative transfer calculations including PRD. It treats line blending  self-consistently.
We use the original RH 1D version to study the basic formation mechanism in Section~\ref{sec:steady} with the FALC \citep{FALC} atmosphere. 
 
For the 3D atmosphere, we use the 1.5D version of RH \citep{Pereira:2015aa}, which treats each column in a  3D atmosphere as an independent plane-parallel atmosphere.
The advantage of this version of RH over the original version is that the code is parallelized
using the Message Passing Interface (MPI) such that the different columns can be calculated
simultaneously. To save computational effort, we do not solve for hydrogen and carbon simultaneously but solve first for hydrogen 
and store the radiation fields to be used for the carbon solution. 

In order to do so accurately, we modified RH so that it stores the ratio between the emissivity and absorption coefficients \citep[$g_{ij}$ in Eq. (26) of][]{RH-uitenbroek01} for Ly$\alpha$ in the hydrogen calculation. The code then reads these ratios in the computation for carbon and so includes the correct radiation field including PRD effects for the radiative pumping by  Ly$\alpha$  in the \ci\ continuum (see Sec.~\ref{sec:steady}).

For the hydrogen computation, we also modified RH to maintain the non-equilibrium hydrogen ionization degree as present in the 3D model atmosphere (see Sec.~\ref{subsec:modelatmos}).  We keep the proton density fixed and remove the rate equation for the continuum to keep the rate equations internally consistent. This method is a fairly accurate approximation of the full non-equilibrium non-LTE radiative transfer because the timescales for the ionization balance are long, but the excited levels of hydrogen adapt very fast and can thus be treated assuming statistical equilibrium. For further details, we refer the reader to
\citet{2017A&A...597A.102G},
who discuss this method in the context of helium.

In Sec.~\ref{sec:diag} we also make use of non-LTE calculations of the \oil\ presented in \citet{hhlinoi}.

 \subsection{Model atmospheres} \label{subsec:modelatmos}
In this study, we use two model atmospheres.
In Section~\ref{sec:steady}, we use the  semi-empirical solar chromosphere model FALC \citep{FALC} to study the basic formation mechanism. 
The FALC model is a semi-empirical atmosphere model that describes the atmosphere in an averaged fashion without dynamics, hence is suitable for studying the line formation qualitatively. 
However, the nature of the solar chromosphere is far from static. To study the radiation from such a dynamic system, we need a more realistic atmosphere model than FALC. We chose to use a model computed with the RMHD code Bifrost \citep{Bifrost-code11}.
In Section~\ref{sec:move},  we use a snapshot from the publicly available 3D simulation {\tt en024048\_hion}
\citep{cb24bih}. 
The snapshot extends 24x24x16.8 Mm in physical space, mapped onto a grid of 504x504x496 cells.
It covers the upper convection zone, the chromosphere, the transition region and the lower corona and includes 
a large-scale magnetic field with two polarities separated by about 8\,Mm with an average unsigned magnetic field strength in 
the photosphere of 50\,G.  
This particular snapshot has also been used in the study of the line formations of H$\alpha$ \citep{jlha12}, 
\mgii\ h and k \citep{mg1,mg2,mg3},  \cii\ 133.5 nm multiplet \citep{bhavna1, bhavna2}, and the \oil\ \citep{hhlinoi}. We refer the reader to \citet{cb24bih} for more details
on this simulation. \edt{We stress here that the purpose of the current work is to investigate the diagnostic potential of the \cil\ and not
to test the realism of the atmospheric model. It is well known that the employed Bifrost model fails to reproduce observations in
several aspects, \eg, chromospheric lines are too narrow in the model on average and several emission lines are too weak. These 
discrepancies may be due to a lack of numerical resolution in the models or a lack of physical processes like ambipolar diffusion
\citep[\eg,][]{Martinez2017}. However, for our purposes, it is enough that the model atmosphere contains the proper parameter
range (there are both wide lines and strong emission at locations in the simulation but not enough to reproduce the mean profiles).}

\subsection{Quintessential Model Atom}
\begin{figure}
   \centering
    \includegraphics[width= \columnwidth]{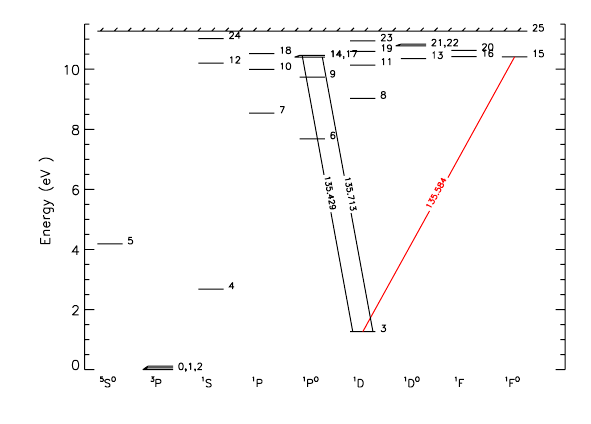}

   \caption{Term diagram of the 26-level \ci\ model atom we use in this study. 
 The energies and designations of each numbered level are listed in Table \ref{tb:ci_terms}. 
 The energies of the levels of the term $2s^2\,2p^2\,^3\!P$ (levels 0,1,2) are very close to each other 
  and the energy differences within the term are exaggerated in this diagram. The same is true for energy levels 14,17 and 21,22.
  Only the transitions within the IRIS wavelength range are shown in the diagram with vacuum wavelengths in nm and the transition of special interest here, the \cil, in red.
   }
   \label{fig:c_term_diag}
\end{figure}

We start with a model atom containing 168 levels with atomic data taken from the 
HAOS-Diper package \citep{diper94}. We do not include hydrogen collisions since atomic data for these are highly uncertain and the
study by \citet{fabbianci} indicates that they are not important. We simplify this model atom for our purposes. In the simplification procedure,
we make sure that at each step we get the same intensity from the FALC model atmosphere 
in the \cil\ but also in two other \ci\ lines in the IRIS wavelength range at 135.43\,nm and at 135.71\,nm.
Since the lines we focus on here are within the singlet system and do not have any direct coupling with the triplet system, we 
start by excluding all states in the triplet system except the ground states. We do still include the radiative recombination rates into the triplet system with
the approach given in \citet{hhlinoi}. 
The corresponding effective recombination coefficients are listed in Table \ref{tb:cr}.  
We then further simplify our model atom by merging levels, following the procedure described in \cite{Bard08}.  
The term diagram  of the final 26-level model atom is shown in Fig. \ref{fig:c_term_diag} and the corresponding atomic parameters are listed in Table \ref{tb:ci_terms}. 
The word "merged" in the level designation means that the level is such a merged super level. The level designation comes from the level with the lowest energy among the merged levels.

\begin{table*}
\caption{Effective recombination coefficient from the continuum to $ 2s^2 2p^2 \, ^3\!P_1$ through the triplet states. } 
\begin{center}
\begin{tabular}{l*{9}{c}r}
\hline
&\multicolumn{8}{c}{Temperature (K)}\\
         & 4500&5160&6370&7970&9983&20420&41180 &60170&100000 \\     
\hline\hline
$R^*_{c1}$ ($10^{-19} \ \mathrm{m}^3 \, \mathrm{s}^{-1}$) &1.30167&1.24998&1.16755&1.08249&1.0130&1.25747&3.99362&7.77096&13.0163 \\

\hline\hline
\end{tabular}
\label{tb:cr}
\end{center}
\end{table*}

\begin{table}

\caption{Atomic parameters for \ci.}\label{tb:ci_terms}
\begin{tabular}{rrl}
\hline\hline
Level             & Energy [$cm^{-1}$]& Designation \\
\hline
0                 &0.000 & \ci\ $ 2s^2 2p^2\, ^3\!P_0$ \\
1           &16.400  & \ci\ $ 2s^2 2p^2\, ^3\!P_1$\\
2          &43.400 & \ci\ $ 2s^2 2p^2\, ^3\!P_2$\\
3          &10192.630& \ci\ $ 2s^2 2p^2\, ^1\!D_2$\\
4          &21648.010& \ci\  $ 2s^2 2p^2\, ^1\!S_0$ \\
5          &33735.200& \ci\  $ 2s 2p^3\, ^5\!S_2^o$ \\
6          &61981.820& \ci\  $ 2s^2 2p 3s\, ^1\!P_1^o$ \\
7          &68856.330& \ci\  $ 2s^2 2p 3p\, ^1\!P_1$ \\
8          &72838.251 & \ci\  $ 2s^2 2p 3p\, ^1\!D_2$-merged \\
9          &78532.462& \ci\  $ 2s^2 2p 4s\, ^1\!P_1^o$-merged \\
10        &80562.850 & \ci\  $ 2s^2 2p 4p\, ^1\!P_1$ \\
11          &81769.790& \ci\  $ 2s^2 2p 4p\, ^1\!D_2$ \\
12          &82251.710& \ci\  $ 2s^2 2p 4p\, ^1\!S_0$ \\
13          &83497.620& \ci\  $ 2s^2 2p 4d\, ^1\!D_2^o$ \\
14          &83877.310& \ci\  $ 2s^2 2p 5s\, ^1\!P_1^o$ \\
15          &83947.430& \ci\  $ 2s^2 2p 4d\, ^1\!F_3^o$ \\
16          &83997.906& \ci\  $ 2s^2 2p 4f\, ^1\!F_3$-merged \\
17          &84032.150 & \ci\  $ 2s^2 2p 4d\, ^1\!P_1^o$ \\
18          &84851.530& \ci\  $ 2s^2 2p 5p\, ^1\!P_1$ \\
19          &85399.810& \ci\  $ 2s^2 2p 5p\, ^1\!D_2$ \\
20          &85717.731 & \ci\  $ 2s^2 2p 5f\, ^1\!F_3$-merged \\
21          &86909.475& \ci\  $ 2s^2 2p 5d\, ^1\!D_2^o$-merged \\
22          &87083.365& \ci\  $ 2s^2 2p 3d\, ^1\!D_2^o$-merged \\
23          &88260.370 & \ci\  $ 2s^2 2p 7p\, ^1\!D_2$ \\
24          &88856.030 & \ci\  $ 2s^2 2p 7p\, ^1\!S_0$-merged \\
25          &90859.558& \cii\  $ 2s^2 2p\, ^2\!P_0$ \\
\hline\hline
\end{tabular}
\end{table}

\section{Basic formation mechanism}\label{sec:basicform}

\subsection{Ionization chain and the line formation}\label{sec:steady}

The main ionization process of \ci\ is through photoionization from  the $2s^2 2p^2\, ^3\!P_1$,  $2s^2 2p^2\, ^3\!P_2$ and $2s^2 2p^2\, ^1\!D_2$ states with photoionization from the $2s^2 2p^2\, ^{1}\!D_2$ state being dominant.
$2s^2 2p^2\, ^{1}\!D_2$ is also the lower state of the \cil\, and it is mainly populated from the ground term $2s^2 2p^2\, ^{3}\!P$ by collisions.  
The \cii\ ground state is depopulated through radiative recombination to all excited states, followed by  radiative cascading.

The upper state of the \cil, $2s^2 2p 4d\, ^1\!F_3^o$, is populated by radiative recombination from the continuum, and depopulated mostly to the state $ 2s^2 2p 4p\, ^1\!D_2$. The \cil\ ($2s^2 2p 4d\, ^1\!F_3^o$ - $2s^2 2p^2\,  ^1\!D_2 $), is, however, not the dominant channel of depopulating $2s^2 2p 4d\,  ^1\!F_3^o$.

\begin{figure}
   \centering
   \includegraphics[width= \columnwidth]{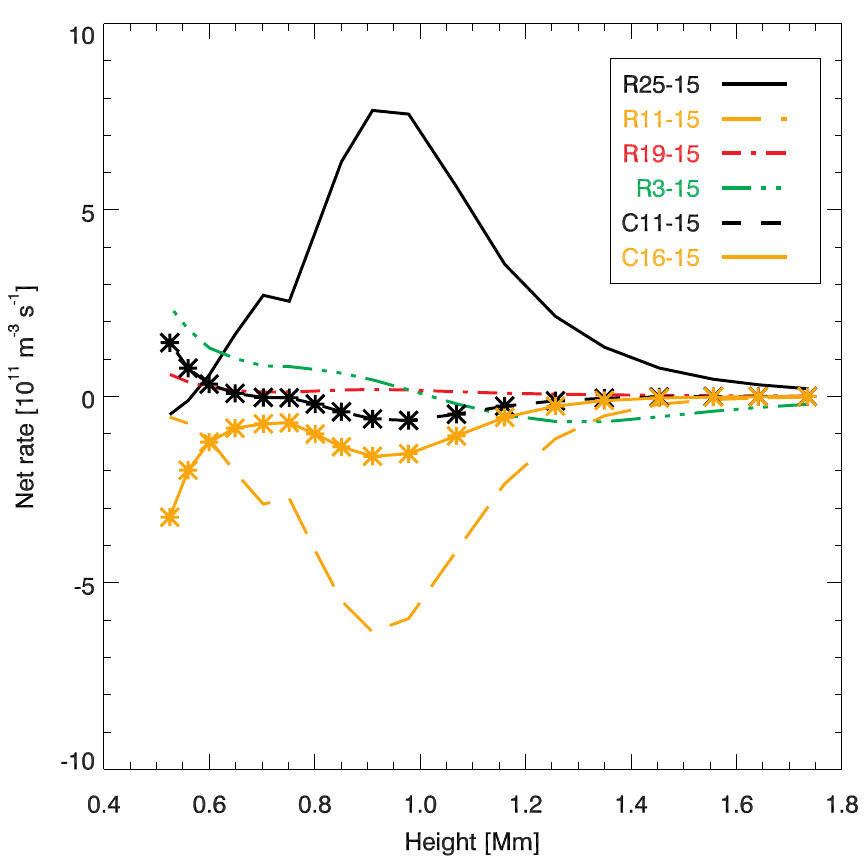} 
  
   \caption{Net rates into the upper level of the \cil, $2s^2 2p 4d\, ^1\!F_3^o$ (level 15  in the \ci\ model atom shown in Figure~\ref{fig:c_term_diag}) in the FALC atmosphere. 
   Positive values denote a net rate into the level, negative values denote a net rate out of the level. 
   Letters R and C in the label stands for radiative net rate and collisional net rate, respectively. 
   The collisional net rates are also overlaid with star symbols.   
   The dominant channel into the $2s^2 2p 4d\, ^1\!F_3^o$ level is from radiative recombination directly from the continuum 
   (level 25, solid black). The main channel out of level 15 is to the level $ 2s^2 2p 4p\, ^1\!D_2$, (level 11, dashed yellow). 
   The net rate corresponding to the 135.58 nm line, denoted as $R15-3$, (green, dashed-dotted-dotted-dotted), is upward, populating
   level 15 (positive) below 1\,Mm height and is downward (negative) above this height.
   Collisions to and from the states $ 2s^2 2p 4p\, ^1\!D_2$ (level 11) and $ 2s^2 2p 4f\, ^1\!F_3$-merged (level 16) also play a role.     }
   \label{fig:nr_15}
\end{figure}

\begin{figure}
   \centering
	 \includegraphics[width=\columnwidth]{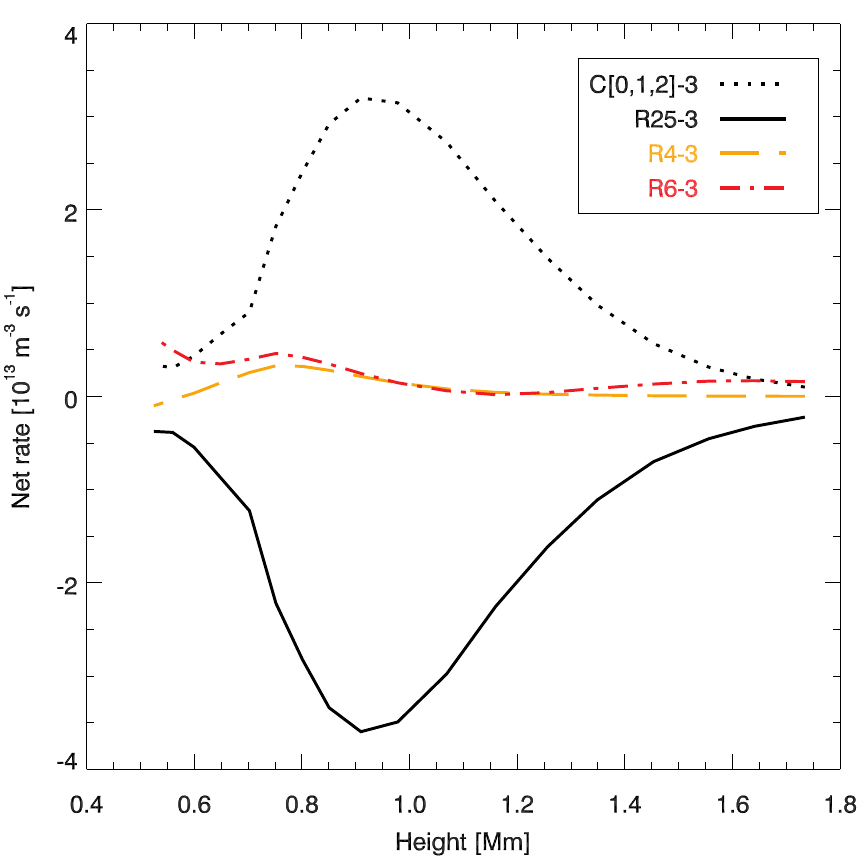}
 \caption{Net rates into the  lower level of the \cil, $2s^2 2p^2\, ^{1}\!D_2$ (level 3 in the \ci\ model atom, see Figure~\ref{fig:c_term_diag}). 
 This level is populated (positive net rate) mostly from the ground term, $2s^2 2p^2\,  ^{3}\!P$ by collisions (dotted black). 
 The dominant depopulation channel (negative net rate) is photoionization (solid black) with some return flow through 
 recombination to excited levels followed by cascading ending up through levels 4 and 6 (denoted R4-3, R6-3 here, 
 dashed brown and dotted-dashed red, respectively).}  
   \label{fig:nr_3}
\end{figure}

The net rates for the $2s^2 2p 4d\, ^1\!F_3^o$ and $2s^2 2p^2\,  ^1\!D_2 $ levels are shown in Fig.~\ref{fig:nr_15} and Fig.~\ref{fig:nr_3}, respectively.

\begin{figure} 
   \centering
  \includegraphics[width= \columnwidth]{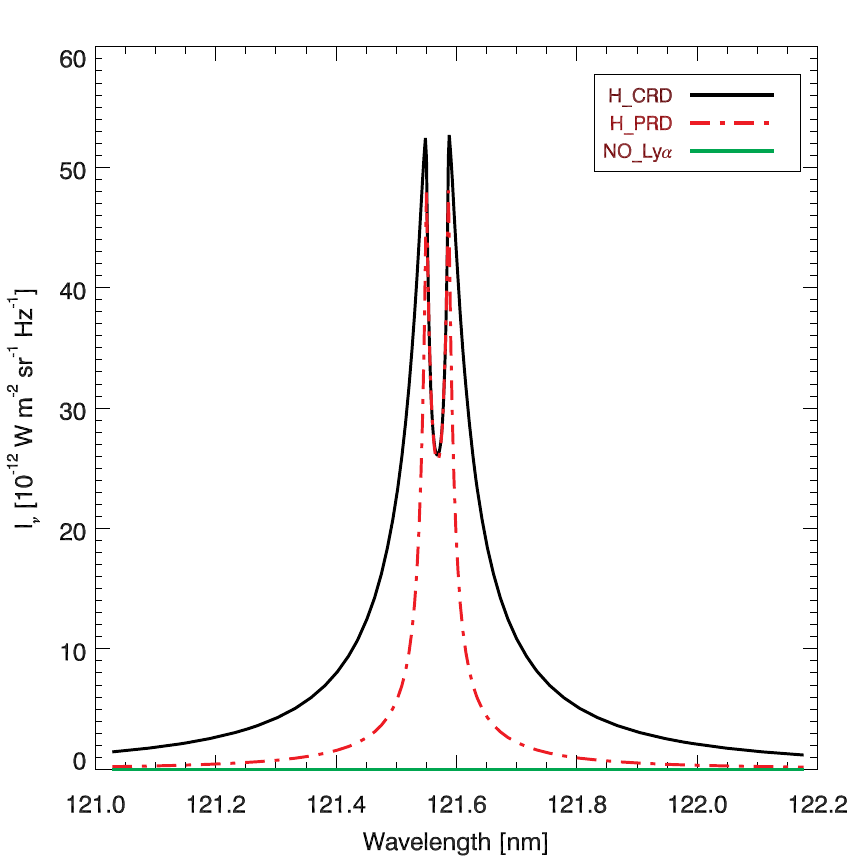} 
  \caption{Vertically emergent intensity from the FALC model with different treatments of the Ly$\alpha$ line. Ly$\alpha$  in CRD (black), in PRD (red dash-dotted), and without the Ly$\alpha$ (green).
   }
   \label{fig:lalb_falc}
\end{figure}

The net-rate analysis shows that the ionization degree of neutral carbon is set by photoionization/radiative recombination and is
therefore sensitive to the radiation field. 
Photoionization from the $2s^2 2p^2\, ^{1}\!D_2$ level requires photons with wavelength below 123.96 nm. This photoionization edge
is just long-ward of the hydrogen  Ly$\alpha$ line at 121.2 nm and the photoionization rate is therefore sensitive to the Ly$\alpha$  intensity
 \citep{semiflareupper}. Although the Ly$\alpha$ intensity is highest close to the line core, for the integrated photoionization rate, its wide
 wings are important. Therefore, the treatment of the frequency redistribution of photons in Ly$\alpha$ is important. Assuming complete frequency redistribution (CRD)
 leads to an overestimate of the wing intensities, and therefore also of the photoionization rate, compared with the proper treatment with 
 partial frequency redistribution (PRD, see Fig.~\ref{fig:lalb_falc}.) 
 
\begin{figure}[htbp] 
   \centering
   \includegraphics[width=\columnwidth]{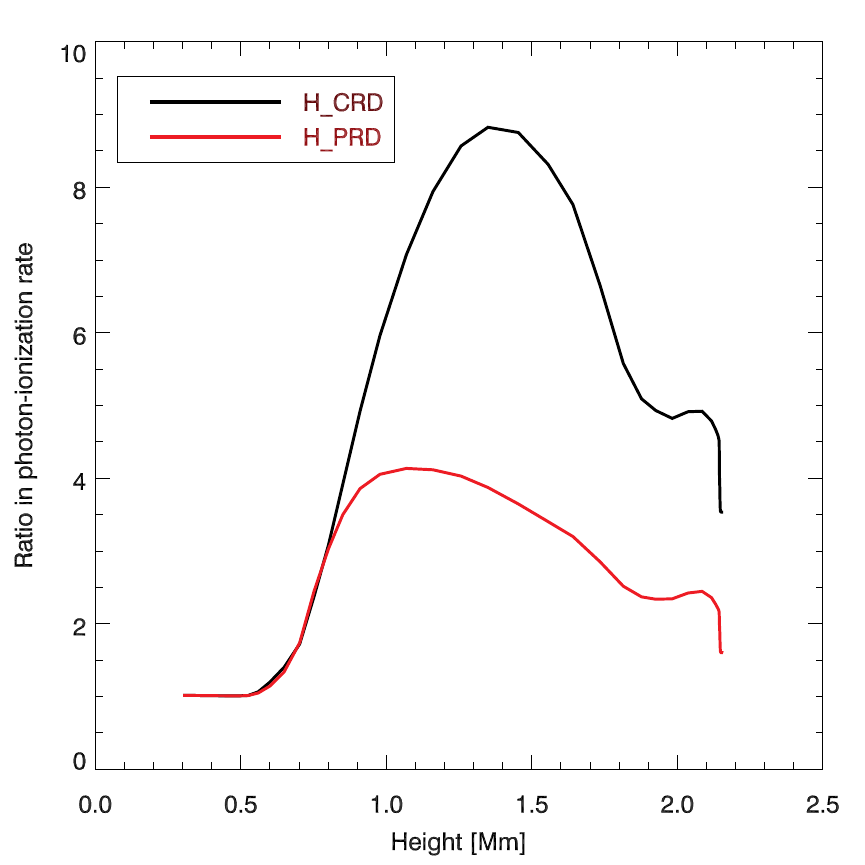} 
   \caption{Ratio of the \ci\ photoionization rate in the FALC model atmosphere with the Ly$\alpha$ line profiles treated with CRD (black) and PRD (red) with respect to the case without Ly$\alpha$ line emission.  
   The definition of the ratio is given in Eq. (\ref{eq:ratio}).}
   \label{fig:prate}
\end{figure}

 We quantify this by defining the following ratio $R$ of photoionization rates:
 \begin{equation}\label{eq:ratio}
R = \frac{\int_{\nu_0}^{\infty} \frac{\sigma_{3 c}(\nu)}{h\nu} J_{\nu} (\mathrm{PRD/CRD})\, \dd\nu}{\int_{\nu_0}^{\infty} \frac{\sigma_{3 c}(\nu)}{h\nu} J_{\nu} (\mathrm{No\ Ly}\alpha)\, \dd\nu}
\end{equation}
where $ \sigma_{3 c}$ is the photoionization cross-section from the $2s^2 2p^2 \ ^{1}D_2$ level to the continuum,
$ J_{\nu} (\text{CRD/PRD})$ is the radiation field with hydrogen lines in CRD/PRD respectively, and $ J_{\nu} (\text{No\_Ly}\alpha)$ is the radiation field without Ly$\alpha$ emission.
This ratio thus quantifies the effect of the Ly$\alpha$ line on the photoionization of \ci. We show $R$ as a function of height in the FALC model in Fig.~\ref{fig:prate}.
It is unity in the deep atmosphere but becomes larger  above 0.5 Mm height. With the assumption of CRD in the Ly$\alpha$ line, the ratio becomes much larger than when assuming PRD owing to the extended Ly$\alpha$ wings in CRD.

\begin{figure}[htbp] 
   \centering
   \includegraphics[width=\columnwidth]{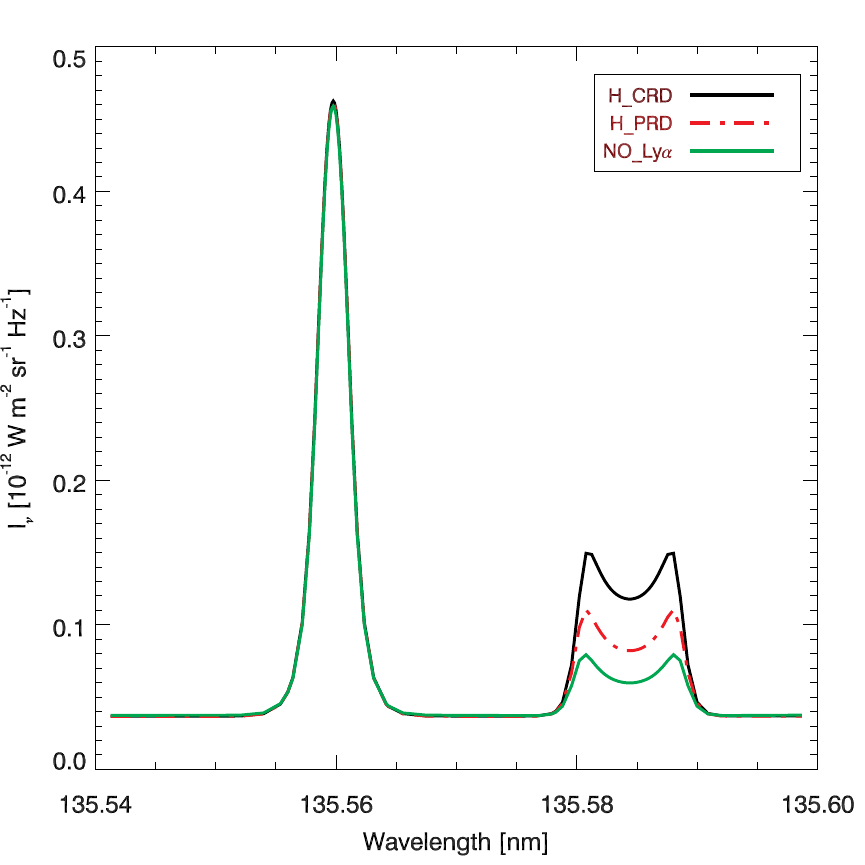} 
   \caption{Resulting \oi\ and \ci\ line intensity with different treatments of the Ly$\alpha$ line. Black:  Ly$\alpha$ in CRD; red dotted-dashed:  Ly$\alpha$ in PRD; green: no  Ly$\alpha$ line.}
   \label{fig:oc_falc}
\end{figure}

Finally, we compare the \ion{O}{1} 135.56 nm and the \ion{C}{1} 135.58 nm lines that are calculated based on treating  Ly$\alpha$ in PRD, CRD, and excluding the line altogether in Fig. \ref{fig:oc_falc}. There are substantial differences in the resulting \cil. The stronger the \ci\ photoionization the stronger the line, which is consistent with the result that the upper level of the line is populated mainly through photorecombinations from \cii.

\begin{figure}[htbp] 
   \centering
    \includegraphics[width= \columnwidth]{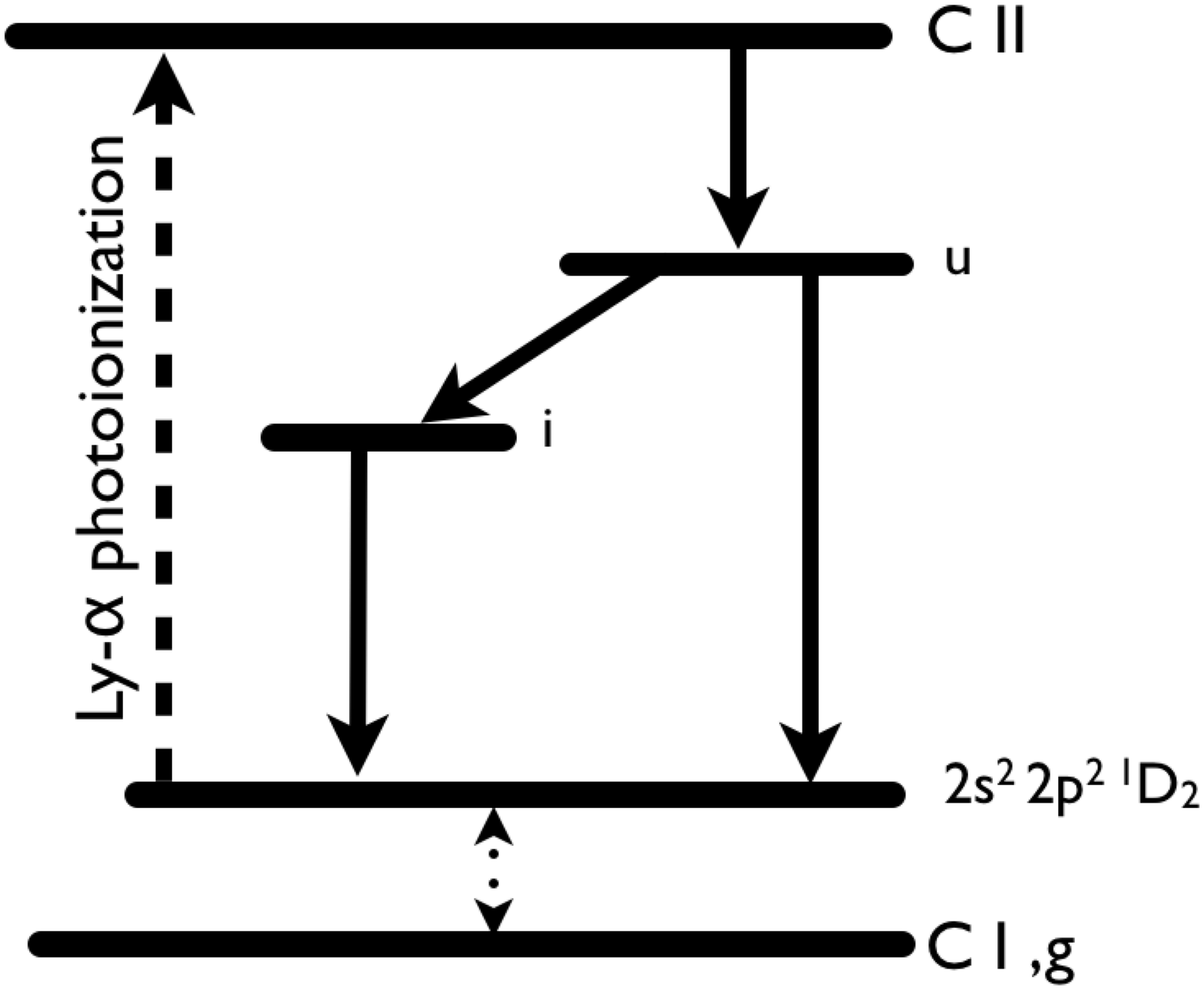} 
   \caption{Schematic C I term diagram that summarizes the basic formation mechanism of the \cil.
   The solid lines represent radiative cascading, the dashed line photoionization, and the dotted line stands for collisions. 
   The ionization degree of C I is set by two steps: first a collisional coupling from the ground term to  $2s^2 2p^2\, ^{1}\!D_2$ and then photoionization by the radiation field blueward of 123.96 nm, including an important contribution from hydrogen Ly$\alpha$ emission. 
   The upper level of the 135.58 nm line, marked with "u", is populated by radiative recombination. The downward rate from "u" channel through intermediate states, denoted as "i", or directly back to $2s^2 2p^2\, ^{1}\!D_2$.   }
   \label{fig:c_term}
\end{figure}

To summarize the basic formation mechanism of the \cil, we show the dominant channels of the \ci\ formation in Fig.~\ref{fig:c_term}. 

\subsection{Line formation in a dynamic atmosphere}\label{sec:move}

So far, we have studied the basic line-formation mechanisms using the 1D, static, semi-empirical model atmosphere FALC.
We now turn to how the \cil\ is formed in a more realistic atmosphere with velocity fields. For this purpose, we use the
Bifrost simulation {\tt en024048\_hion} \citep{cb24bih}. To illustrate the line formation, we use the four-panel diagrams
introduced by \citet{4panel}.

The vertically emergent intensity in a 1D plane-parallel semi-infinite atmosphere can be written as 

\begin{equation*}
I_\nu = \int_{0}^{\infty}  S_{\nu}(z) \, e^{-\tau_{\nu}(z)} \, \chi_{\nu}(z) \, \dd z.
\end{equation*}

where 
$S_{\nu}$, $\chi_{\nu}$, and $\tau_{\nu}$ are the source function, opacity, and optical depth, respectively.

In this equation, the integrand describes the local creation of photons ($S_{\nu}(z)\chi_{\nu}(z) \dd z$) and the fraction of the those that escape ($e^{-\tau_{\nu}(z)}$), and the integrand is thus a natural definition of the contribution function to intensity on a height scale.

\begin{equation*}
C_{I_{\nu}}(z) = S_{\nu}(z) e^{-\tau_{\nu}(z)} \chi_{\nu}(z).
\end{equation*}

We can rewrite this contribution function as

\begin{equation*}
C_{I_{\nu}}(z) = S_{\nu}(z) \tau_{\nu}(z) e^{-\tau_{\nu}(z)} \frac{\chi_{\nu}(z)}{\tau_{\nu}(z)}
\end{equation*}

where the term $\tau_{\nu}(z) e^{-\tau_{\nu}(z)}$ peaking at $\tau_{\nu}=1$ represents the Eddington-Barbier part of the contribution function, $S_{\nu}(z)$ gives the source function contribution and 
${\chi_{\nu}(z)}/{\tau_{\nu}(z)}$ picks out effects of velocity gradients in the atmosphere \citep[see][]{4panel}.
 
\begin{figure}
   \centering
    \includegraphics[width=\columnwidth]{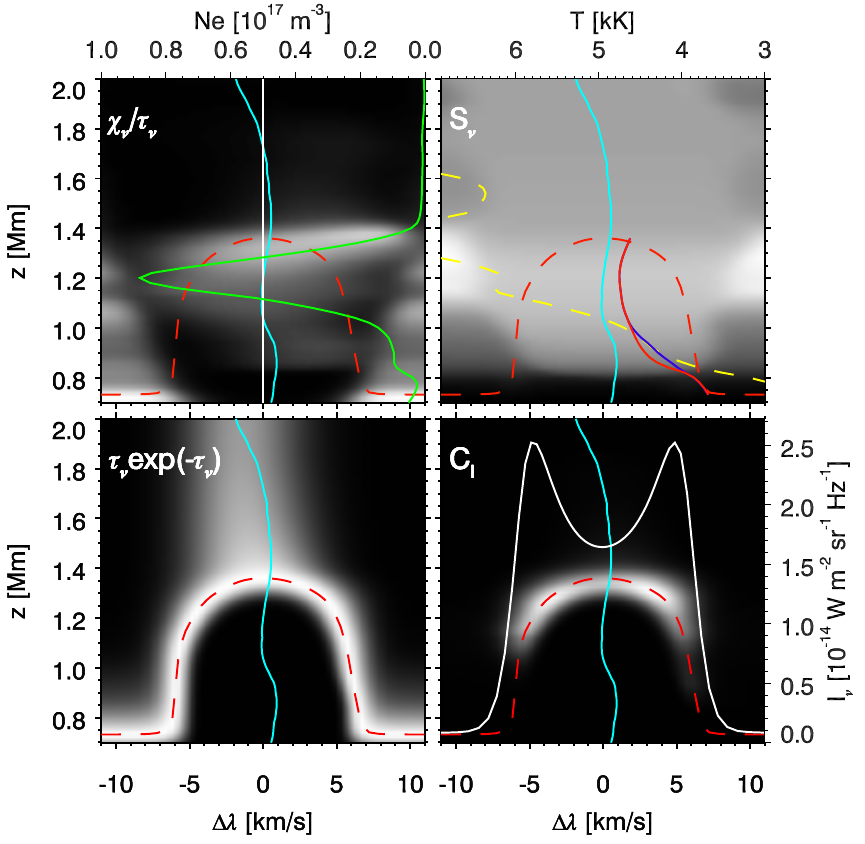} 
     \caption{Intensity formation of the \cil\ in a column of the 3D atmosphere characterized by moderate velocities and 
     almost zero velocity gradients.  
     The label in the top-left corner of each panel shows the quantity shown as a grayscale image as a function of 
     doppler shift from the rest wavelength (positive is redshift) and height $z$.  
     The $\tau_\nu=1$ height is outlined as a red-dashed line and the vertical velocity as a turquoise solid line in all four panels. 
     The white straight line in the upper left panel shows the rest wavelength, and the green solid line gives the electron 
     density with a scale at the top.
     In the panel of $S_\nu$ (top right), the dashed yellow curve denotes the temperature as function of height $z$, 
     while the solid curves denote the value of the source function at a given height at the wavelength that has $\tau_\nu=1$ at
     that height on the red-side (red line), and blue-side (blue line) of the  $\tau_\nu=1$ maximum.
     They are given in radiation temperature units with the scale given at the top.
     In the lower right panel, we show the full contribution function (the product of the terms shown 
     as grayscale images in the other three panels) together with the emergent intensity profile
     with a scale to the right. 
     The line profile is symmetric with two emission peaks and a central depression. 
     These emission peaks are formed where there is a local maximum in the source function and the central depression 
     is caused by the decreasing source function above this height.
 }
   \label{fig:ci_form1}
\end{figure}

\begin{figure}[htbp] 
   \centering
     \includegraphics[width=\columnwidth]{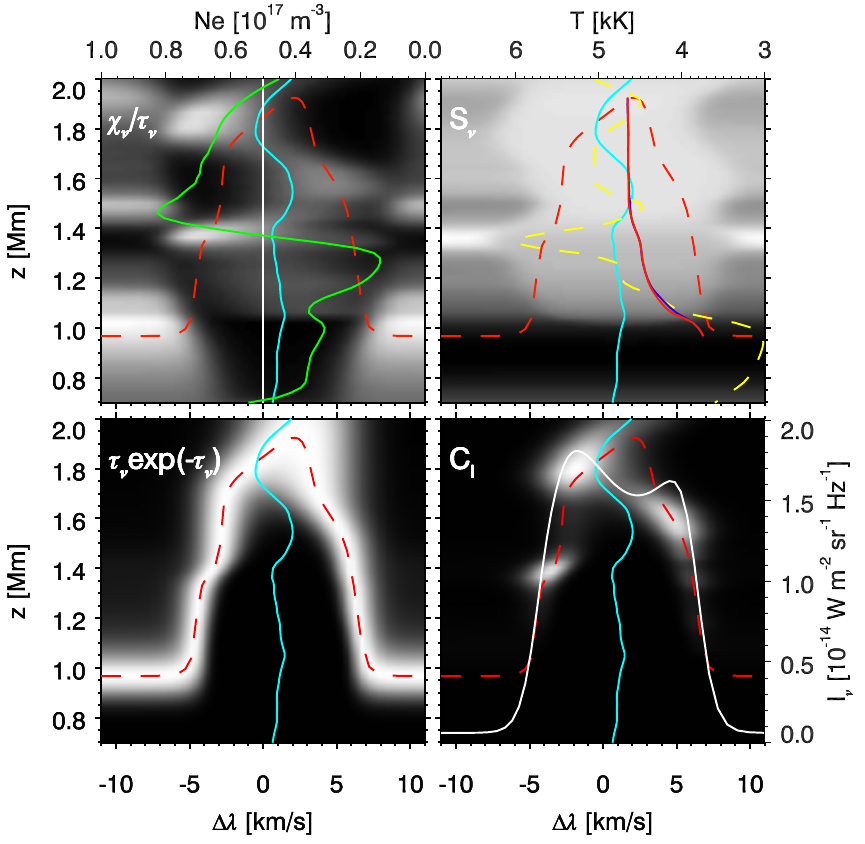} 
     \caption{As Fig.~\ref{fig:ci_form1} but for a column of the 3D atmosphere characterized by moderate velocities and 
     moderate velocity gradients. 
     The intensity profile has two peaks, as is the case in Fig.~\ref{fig:ci_form1}, but the cause is quite different:
     The source function is rather constant with height and the two maxima of the contribution function are caused
     by velocity gradients (the term that is shown in the upper left panel) rather than by a local maximum of the source function.
     }
   \label{fig:ci_form2}
\end{figure}

\begin{figure}[htbp] 
   \centering
     \includegraphics[width=\columnwidth]{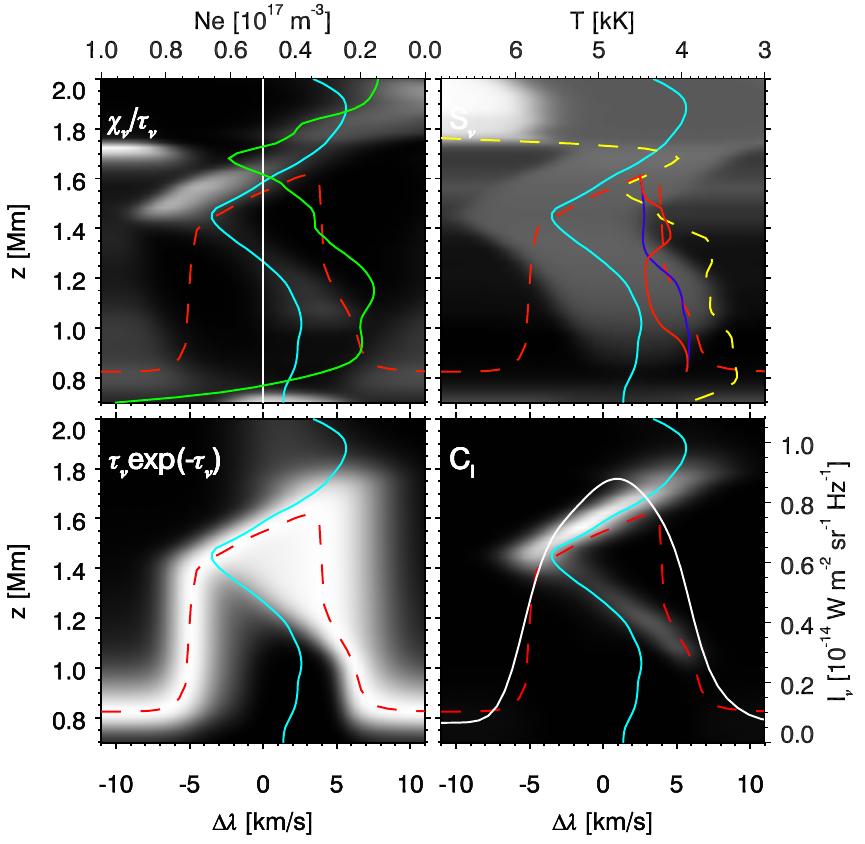} 
     \caption{Same as Fig.~\ref{fig:ci_form1} but for a column of the 3D atmosphere characterized by strong velocities and strong velocity gradients. 
     The formation of the line is driven by the velocity gradient and we do not observe a decreasing source function with height. 
     This results in a singly peaked line profile.  
     }
   \label{fig:ci_form3}
\end{figure}

Fig.~\ref{fig:ci_form1} shows the intensity formation of the \cil\ in an atmospheric column with moderate velocities and almost no 
velocity gradients. The intensity is formed close to $\tau_\nu = 1$ as indicated by the maximum of the contribution function following the $\tau_\nu = 1$ curve. The continuum is formed 
at a height of 0.73~Mm, where the
temperature is 3\,kK but with a decoupled source function at a radiation temperature of 
3.7\,kK. The line core has
optical depth unity  at $z = 1.35$\,Mm.
The upper right panel shows that the source function increases up to $z=1.2$~Mm and decreases higher up. The
local maximum of the source function is the reason for the two emission peaks and the central reversal.
 
Fig.~\ref{fig:ci_form2} shows the intensity formation at another column in the atmosphere where the velocities are larger and there are
significant velocity gradients in the line forming region. The intensity profile (shown in the lower right panel) still has two 
peaks and a central reversal but the reason is different from the case shown in Fig. \ref{fig:ci_form1}. 
The source function (upper right panel) is rather constant with height above $z=1.5$\,Mm, which would give a single peak intensity profile with a flat
top. The reason we still get two emission peaks is that the velocity gradients around the velocity maxima at $z=1.8$\,Mm and
$z=1.6$\,Mm give large $\chi_\nu/\tau_\nu$ terms (upper left panel).

A third example is given in Fig.~\ref{fig:ci_form3}. At this column of the atmosphere, there is a strong velocity gradient above
$z=1.5$\,Mm influencing both the $\tau_\nu=1$ shape with its maximum height at a redshift of 4 \kms\ and the intensity profile.
The source function (the relevant one for the profile blueward of +4 \kms\ is given by the blue line in the upper right panel) is
slowly decreasing above $z=1.4$\,Mm but the $\chi_\nu/\tau_\nu$ term decreases with increasing redshift, so we get a maximum
intensity at a redshift of 1 \kms.

From the selected cases shown above, we learn that the line profile is influenced both by the run of the source function with height
and by velocity gradients in the atmosphere.
 
\section{Diagnostic Potential}\label{sec:diag}

Spectral line-profiles encode information from the conditions in the atmosphere where the photons escape. 
To decode this information, we need to find the line-forming region and then correlate observable quantities with 
atmospheric parameters in this region. We start this section with a discussion of the line-forming region.

\subsection{The line core}\label{sec:lc}

As we have seen in the previous section, the \cil\ has a typical optically thick line-formation with the contribution function
to intensity centered around monochromatic optical depth unity. 
The maximum height we can get information from is at the wavelength where this optical depth unity is the highest. 
This wavelength is not immediately available observationally. In the absence of velocity fields, we have the maximum opacity
at the rest wavelength of the transition, which is also the wavelength of maximum optical-depth-unity height. 
If the source function is monotonically increasing with height, we get an emission line with the maximum intensity at this
wavelength. If there is a local maximum in the source function below the optical-depth-unity height at line center, we get an
emission line with a self-reversal. Dependent on the source function variation with height, we may also get more than two
peaks in the intensity profile.
In the absence of velocity fields, the line center is at the position of the central emission maximum for profiles with an odd number
of peaks and at the position of the central emission minimum for profiles with an even number of peaks. We use this 
wavelength as our observationally defined line-core wavelength in spite of the fact that,
as we have seen in Section~\ref{sec:move}, velocity
gradients may strongly affect the contribution function to intensity and the correlation between this observationally
defined line core wavelength and the  wavelength of maximum $\tau_\nu1$ height.

In the following, we inspect various quantities at this wavelength.

\subsection{The Contribution function}\label{sec:contri}

\begin{figure}[htbp] 
   \centering
   	\includegraphics[width=\columnwidth]{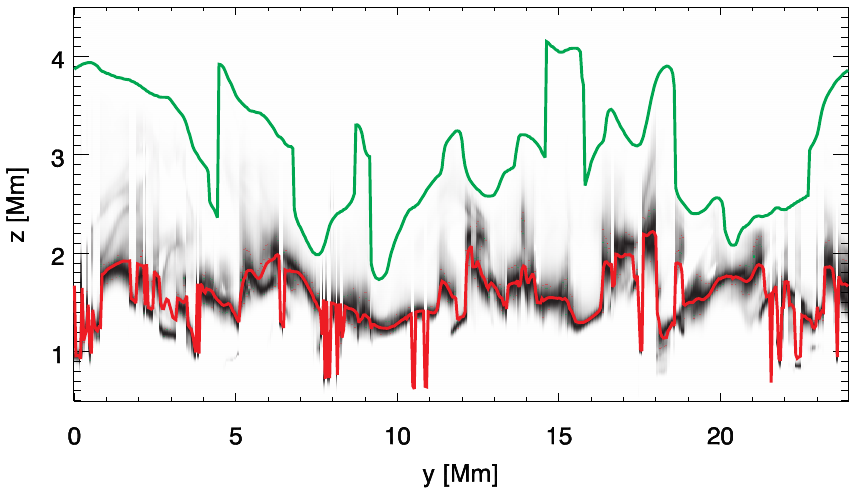} 
     \caption{Contribution function to the intensity at the wavelength of the \cil\ core along the $x=12\,$Mm cut through the
     atmospheric model (Fig. \ref{fig:oici_fh}) shown as a grayscale image. The line core wavelength is defined as the 
     wavelength of the central emission peak for profiles with an odd number of emission peaks and as the wavelength of
     the central absorption for profiles with an even number of emission peaks. Also shown are the $\tau_\nu1$ height for this same wavelength
     (red) and the temperature contour at 30 kK (green).
      The contribution function is normalized at each $y$-value. Note also that the aspect ratio is with an expanded height scale. 
      The maximum of the contribution function closely follows the $\tau_\nu1$ height, consistent with a
     typical optically thick formation.}
   \label{fig:oici_contri}
\end{figure}

In Fig.\ref{fig:oici_contri}  we show the  contribution function to intensity at the line-core wavelength (as defined in Sec.~\ref{sec:lc}) along a selected cut at $x=12$~Mm in Fig. \ref{fig:oici_fh}. 
The red line in Fig.~\ref{fig:oici_contri} represents the $\tau_\nu1$ height. It is clear that the maximum of the contribution
function to intensity is tightly correlated with this height, consistent with an optically thick formation of the line.

\subsection{Formation height}\label{sec:fh}

\begin{figure}
   \centering
   	\includegraphics[width=\columnwidth]{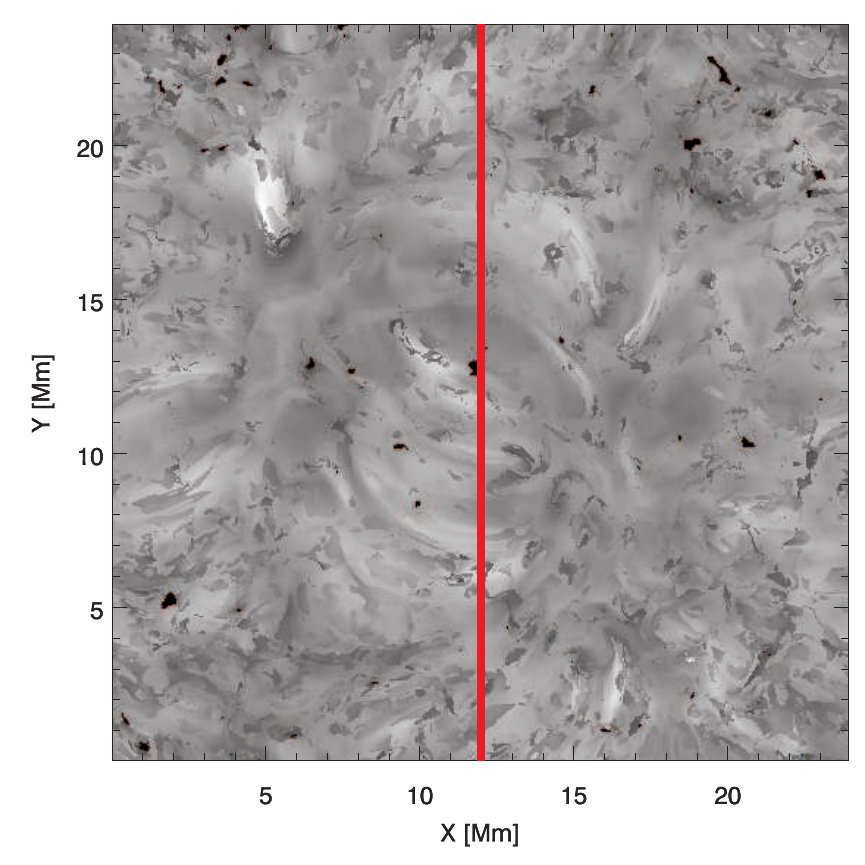} 
     \caption{Formation height of the \ci\ line core. 
     The cut indicated with a red line at $x=12$~Mm gives the 2D cut used in Fig.~\ref{fig:oici_contri}.}
   \label{fig:oici_fh}
\end{figure}

As the C I line has an optically thick formation, the $\tau_\nu=1$ height could be a natural choice for its formation height. 
However, to be consistent with the treatment with the \oil\ \citep{hhlinoi}, we follow the same definition for the \cil\ and use the contribution-function-weighted averages to define the formation height and other quantities:
\begin{equation}\label{eq:vfh}
X_{\rm fh} = {\int_{0}^{\infty} C_{I}(z) X(z) \dd z \over \int_{0}^{\infty} C_{I}(z) \dd z},
\end{equation}
where $X$ is the quantity under consideration and $X_{\rm fh} $ is its  contribution-function-weighted average.

The result for the contribution-function-weighted \cil-formation height is shown in Fig.\ref{fig:oici_fh}. The loop-like features show the influence of the magnetic structures on the line opacity. At some points, the formation height is very much lower than at neighboring points (seen as darker small patches, also visible along the selected cut shown in Fig.~\ref{fig:oici_contri}). These deviating formation heights are associated with the algorithm for selecting the line core wavelength not finding the actual wavelength of the maximum $\tau_\nu=1$ height.

In the further analysis in this section, we use plasma properties calculated using Eq.~\ref{eq:vfh}.

\subsection{Velocity}\label{sec:v}

\begin{figure}
   \centering
   \includegraphics[width=\columnwidth]{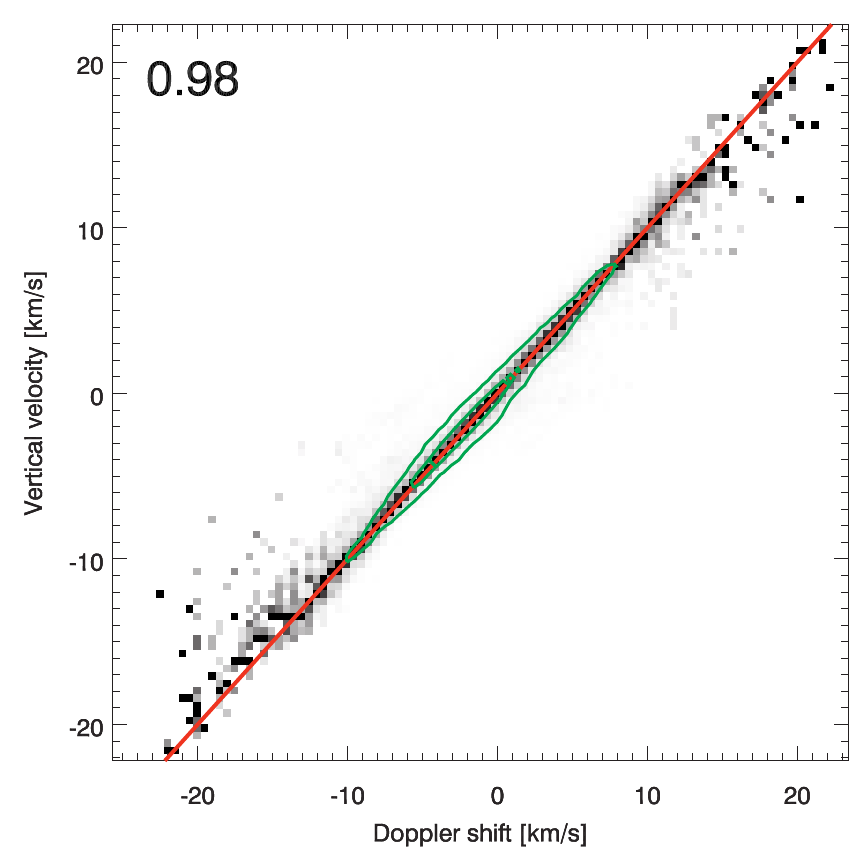}
   \caption{Probability density function (PDF) of the contribution-function-weighted vertical velocity  as function of the Doppler shift of the line-core. 
   The Pearson correlation coefficient is shown in the upper left corner. 
   The green inner contour encloses 50\% of all pixels and the outer contour 90\%.
   The diagonal red straight line denotes $y = x$. 
  The correlation is quite tight, which makes the \cil\ a good velocity diagnostic.}
   \label{fig:vz_ds}
\end{figure}

With the  line-core frequency defined in Sec.~\ref{sec:lc}, we can obtain the Doppler shift with respect to the rest frequency and compare with
 the contribution-function-weighted vertical velocity.
In Fig.~\ref{fig:vz_ds} we find the correlation to be quite tight, making \cil\ a good velocity diagnostic.

\subsection{Electron density}

\begin{figure} 
   \centering
   \includegraphics[width=\columnwidth]{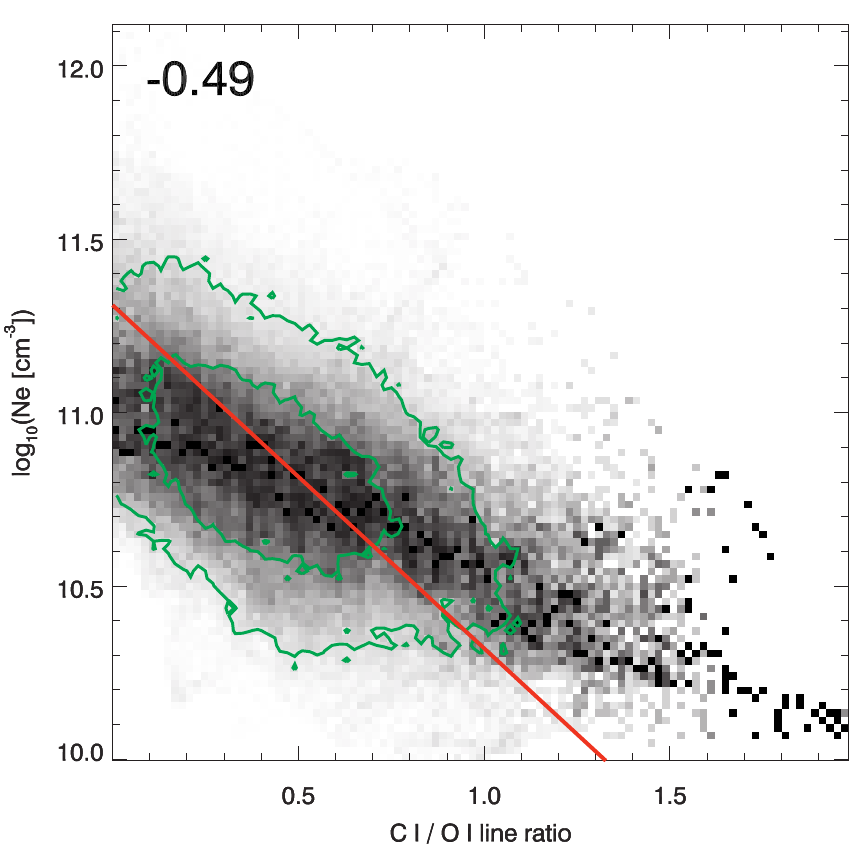}   
        \caption{
        PDF of the electron density as a function of the \ci/\oi\ total line intensity ratio. 
       The Pearson correlation coefficient is shown in the upper left corner. 
        The green inner contour encloses 50\% of all pixels and the outer contour 90\%.
        The red straight line representing $y = -x + 11.31$  indicates $I(\text{C I})/I(\text{O I}) \sim 1/N_\mathrm{e}$.}
   \label{fig:ratio_ne}
\end{figure}

It is interesting to investigate the diagnostic power of the \oil\ and the \cil\ combined, because they are observed together with IRIS.

As discussed in \citet{hhlinoi}, the \oil\ intensity is proportional to the  square of the electron density: the line is formed by recombination from \ion{O}{2} followed by radiative cascading to the upper level of the transition. The recombination rate scales linearly with the electron density. The \ion{O}{2} density is proportional to the proton density (and hence the electron density), which combined yields the quadratic dependence.

In Section~\ref{sec:steady}  we show that the \cil\ formation is driven by  ionization by Ly$\alpha$ photons followed by radiative recombinations. 
Such a process scales linearly with the electron density (through the recombination) but of course will also be dependent on the  Ly$\alpha$ radiation field. As a result, we anticipate that the \ci/\oi\  line ratio scales as 
$\sim J(\mathrm{Ly}\alpha)/N_\mathrm{e}$, where J(Ly$\alpha$) is the radiation field in the Ly$\alpha$ transition. 
In Fig. \ref{fig:ratio_ne} we show the correlation between the electron density at the average formation height  and the \ci/\oi\  line ratio. It shows a roughly linear correlation, but with a considerable spread.

\subsection{Formation height difference}

\begin{figure} 
   \centering
   \includegraphics[width=\columnwidth]{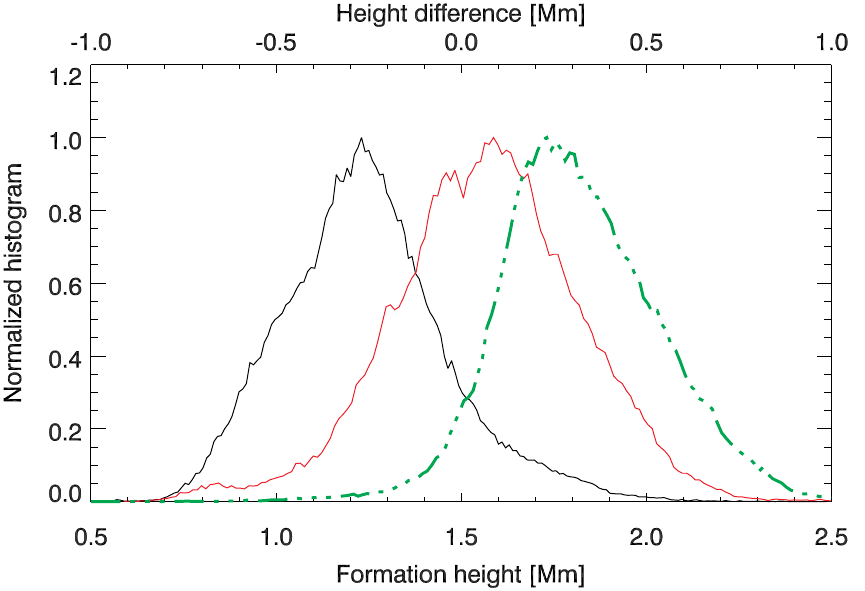}
   \caption{Histograms of the formation height of the \oil\ (black), the \cil\ (red) with the axis below. 
   The histogram of the formation height difference (\ci\ - \oi) is given as a green dash-dotted line with the axis at the top.  The \ci\ line forms mostly higher than the \oi\ line.  }
   \label{fig:histo_fh}
\end{figure}

\begin{figure}
   \centering
   \includegraphics[width=\columnwidth]{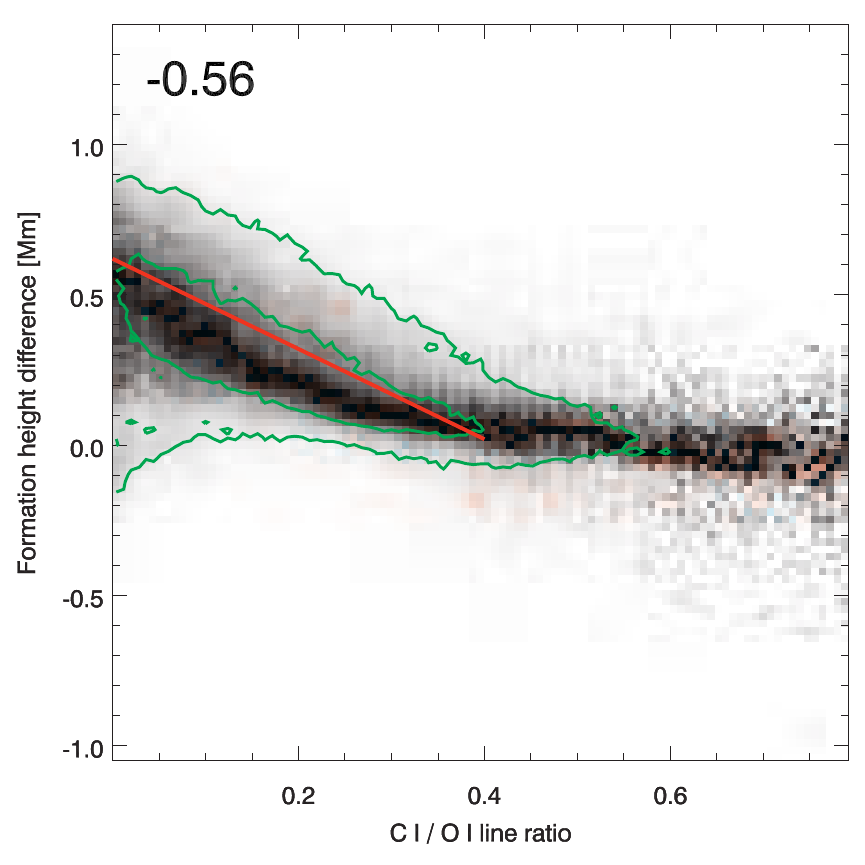}
   \caption{ PDF of the formation height difference  [\ci\ - \oi] as a function of the \ci/\oi\ line-core intensity ratio.
   The Pearson correlation coefficient is shown in the upper left corner. 
   The green inner contour encloses 50\% of all pixels and the outer contour 90\%.
   The red straight line denotes $y = - 1.5x + 0.62$.}
   \label{fig:dh_r}
\end{figure}

From the histogram of the formation heights of the \oil\ and the \cil\ (Fig. \ref{fig:histo_fh}), it is clear that these two lines map slightly different layers in the atmosphere.
In general, the \cil\ forms higher than the \oil. The distribution of the \cil\ peaks around 1.5 Mm,  while the \oil\ distribution peaks around 1.2 Mm.

The two lines have different sensitivity to the electron density (see Fig. \ref{fig:ratio_ne}). This also means that there is a correlation between
the \ci/\oi\ line-core intensity ratio and the formation height difference --- a low intensity ratio corresponds to a high intensity oxygen line 
formed low down and thus with a large formation height difference. This correlation is shown in Fig. \ref{fig:dh_r}. The correlation is clear but the spread is rather large.

\subsection{The velocity difference and the velocity gradient}

\begin{figure}
   \centering
   \includegraphics[width=\columnwidth]{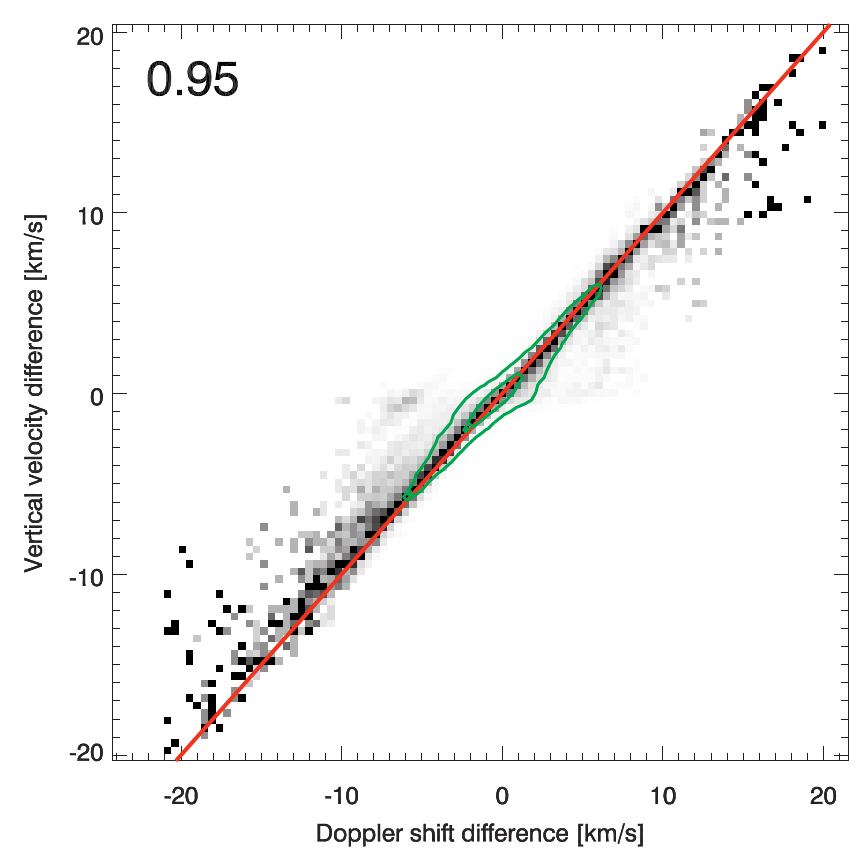}
   \caption{PDF of the difference in correlation-function-weighted vertical velocity between the \ci\ and the \oi\ lines as a function of the Doppler shift difference. 
   The Pearson correlation coefficient is shown in the upper left corner. 
   The green inner contour encloses 50\% of all pixels and the outer contour 90\%.
   The diagonal red straight line denotes $y = x$. }
   \label{fig:vzd_dsd}
\end{figure}

Because the \oil\ and the \cil\ probe slightly different layers in the atmosphere and their Doppler shifts are good velocity indicators, 
it is possible to take the difference between their individual Doppler shifts and measure the velocity difference between these two layers.
In Fig. \ref{fig:vzd_dsd} we show the correlation between the difference between the contribution-function-weighted velocities of the lines and the difference of their individual Doppler shifts. The correlation is strong. Together with the correlation between the line ratio and the formation height difference (Fig.~\ref{fig:dh_r}), it is possible to derive a rough estimate of the velocity gradient in the middle chromosphere.

\section{Discussion and conclusions}\label{sec:concl}

We presented an analysis of the  formation of the \cil\ in the FALC 1D semi-empirical atmosphere model and a 3D RMHD model computed with the Bifrost code. We show that the line has a typical optically thick line formation.

The \ion{C}{1}/\ion{C}{2} ionization balance is mainly driven by photoionization from $2s^2 2p^2 \ ^{1}D_2$, with the radiation field in  Ly$\alpha$ playing a central role. Proper modeling of the \cil\  should therefore incorporate non-LTE modeling of  hydrogen including PRD in the Ly$\alpha$ line. In a static atmosphere, the line core usually appears with a central reversal, due to a local maximum of the source function below 
the maximum $\tau_\nu=1$ height.
However, if velocity gradients are present in the atmosphere then the line profiles can have more different, usually asymmetric, shapes. Interestingly, we find some instances where a strong velocity gradient close to the line-core formation height can cause a single-peaked profile
in spite of a source function decreasing with height at the maximum $\tau_\nu1$ height (which would lead to a central reversal in the
absence of the velocity gradient). We also find cases of the opposite: velocity gradients leading to a central reversal even though the source
function monotonically increases with height. 

The Doppler shift of the line core is a good velocity diagnostic.

The  \cil\ can be combined with the nearby \oil\ for increased diagnostic power. The \cil\ typically forms slightly higher than the \oil, and both are good velocity diagnostics. The difference in their Doppler-shifts is therefore a good indicator of the velocity difference between their formation layers. By combining this with the correlation of the \ci/\oi\ line ratio and the formation height difference one can derive an estimate of the actual velocity gradient. Because the line ratio - height difference correlation is weak and has considerable spread, this estimate is, however, rather coarse.

Furthermore, we find that the  \ci/\oi\  total line intensity ratio is inversely proportional to the electron density: $I(\text{C I})/I(\text{O I}) \sim 1/N_\mathrm{e}$, but again with a large spread in the correlation.

We stress that these relations have been derived from a Bifrost model snapshot that is typical of the quiet Sun. A flare (where the \cil\ is often stronger than the \oil) has very different atmospheric properties and the relations derived here cannot be used under such conditions.

We also note that all correlations are derived from radiative transfer calculations using the 1.5D approximation. The Ly$\alpha$ line is, however, strongly influenced by 3D effects 
\citep{2015ApJ...803...65S,2017A&A...597A..46S},
and this might have an influence on the results presented here.

\newpage

\renewcommand\refname{Reference}
\bibliography{CI}

\end{document}